\documentclass[a4paper]{jpconf}
\usepackage{graphicx}
\usepackage{amssymb}

\begin{document}

\title{ The influence of small scale magnetic field
        on the heating of J0250+5854 polar cap
      }

\author{D.P. Barsukov$^{1,2}$, M.V. Vorontsov$^{2}$, I.K. Morozov$^{1,2}$ }

\address{$^1$ Ioffe Institute, Saint Petersburg, Russia}

\address{$^2$ Peter the Great St. Petersburg Polytechnic University, Saint Petersburg, Russia}

\ead{bars.astro@mail.ioffe.ru}

\begin{abstract}
The influence of surface small scale magnetic field
on the heating of PSR J0250+5854 polar cap
is considered.
It is assumed that polar cap is heated only
by reverse positrons, accelerated in pulsar diode.
It is supposed that pulsar diode is in stationary state
with lower plate nearby the star surface (polar cap model),
occupies all pulsar tube crosssection
and operates in regime of
steady space charge limited electron flow.
The influence of small scale magnetic field on
electric field inside pulsar diode
is taken into account.
To calculate the electron-positron pairs production rate
we take into account only the curvature radiation of
primary electrons and its absorption in magnetic field.
It is assumed that part of electro-positron pairs
may be created in bound state (positronium).
And later such positroniums are photoionized by
thermal photons from star surface.
\end{abstract}

\section{Introduction}
\begin{figure}[h]
\begin{minipage}{14pc}
\includegraphics[width=14pc]{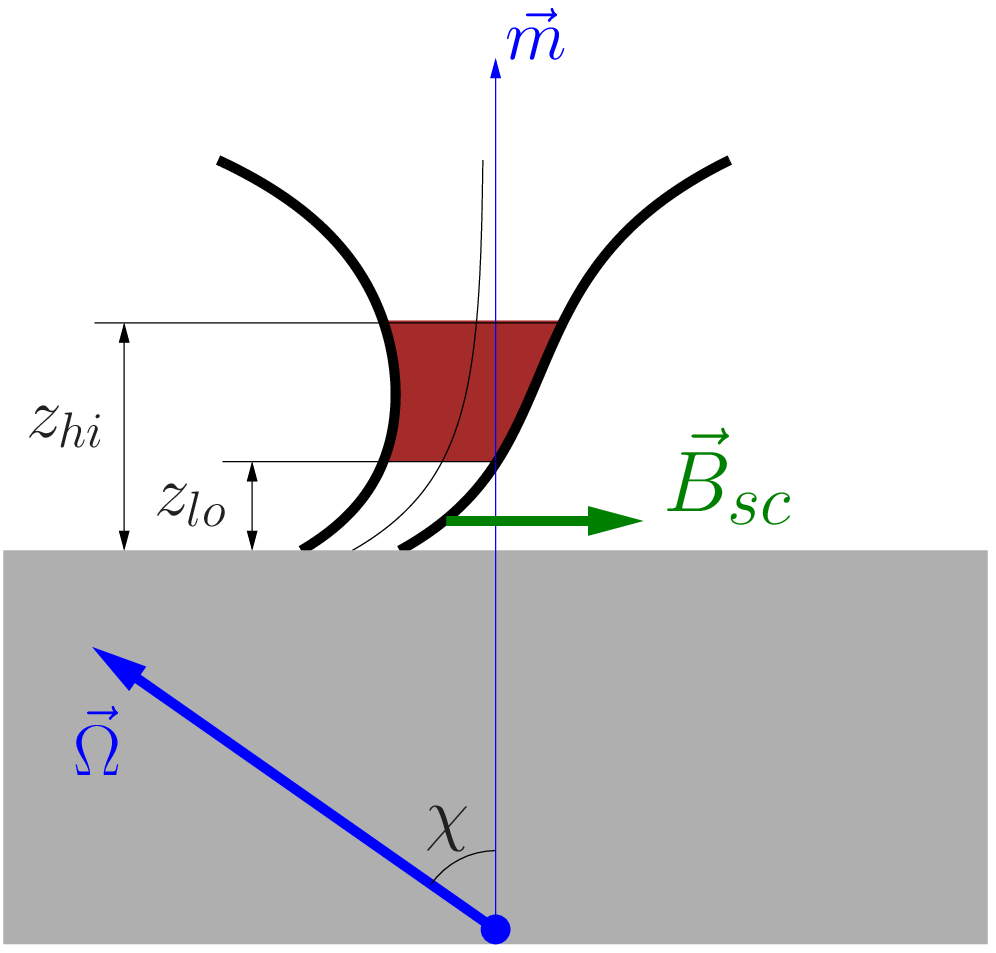}
\caption{\label{fig_diode}
A sketch of the vicinity of an inner gap.
Neutron star is shown by gray area,
boundaries of pulsar tube are shown by black lines,
the inner gap is shown by brown area.
       }
\end{minipage}\hspace{2pc}%
\begin{minipage}{14pc}
\begin{center}
\includegraphics[width=8pc]{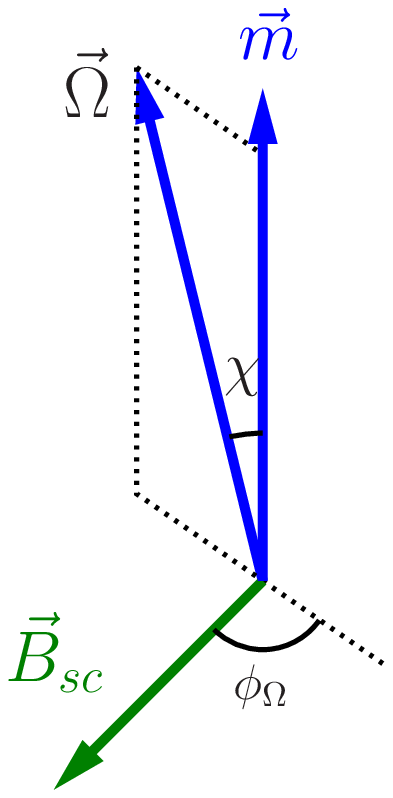}
\end{center}
\caption{\label{fig_angles}The definition of angles  $\chi$ and $\phi_{\Omega}$.
        }
\end{minipage}
\end{figure}

Radiopulsar J0250+5854 rotates with period $P=23.54 \mbox{\ s}$ 
\cite{Tan2018}
and is the slowest pulsar among rotation powered pulsars \cite{ATNF}.
It is old pulsar with spin down age
$\tau=13.7 \cdot 10^{6} \mbox{ years}$,
$\dot{P} = 2.71 \cdot 10^{-14}$,
its spin down energy loss rate $\dot{E}$ is equal to
$\dot{E} = 8.2 \cdot 10^{28} \mbox{erg}/\mbox{s}$, 
the strength $B_{dip} $ of dipolar magnetic field at pole 
estimated by pulsar slowdown is
$B_{dip}=5.1 \cdot 10^{13} \mbox{\ G}$,
distance $D_{DM}$ estimated by dispersion measure is 
$D_{DM}=1.56 \mbox{ kpc}$ \cite{ATNF}. 
Such pulsars lie beyond conventional pulsar "death line"\ , 
see, for example, \cite{HardingMuslimov2002,Kantor2004},
and usually its radio radiation is explained by the presence
of small scale surface magnetic field,
see, for example, \cite{Arons2001,Gil2001,Zhang2002,Muslimov2011}.
It is worth to note that
the radio radiation of J0250+5854 also may be explained 
in case of pure dipolar magnetic field 
if it is take into account that the $B_{dip}$ value is calculated
at assumption aligned pulsar $\chi = 0^{\circ}$
and braking due to magneto-dipolar losses \cite{Beskin},
where $\chi$ is inclination angle i.e.
angle between vector of magnetic dipole momentum $\vec{m}$ 
and vector of angular velocity of star rotation $\vec{\Omega}$,
$\Omega = 2\pi / P$, see fig. \ref{fig_diode} and fig. \ref{fig_angles}.
In case of pulsar braking due to current losses
and nearby orthogonal pulsars $\chi \approx 90^{\circ}$ 
dipolar magnetic field is substantially larger 
than $B_{dip}$ value estimated by slow down 
\cite{BeskinNokhrina}
that moves the pulsar to "life zone"\ \cite{Beskin}.
In this paper we will not consider a such possibility
and assume that the $B_{dip}$ value is 
the right estimation of dipolar magnetic field strength.
In case of large surface magnetic field 
$B_{surf} \gtrsim 4.4 \cdot 10^{12} \mbox{ G}$
electron-positron pairs may be produced 
in bound state (positronium) \cite{UsovMelrose1995}.
The influence of this process on pulsar electrodynamics,
pair generation and polar cap X-ray luminosity of radio pulsars 
has been thoroughly considered in many papers,
see, for example, \cite{Gil2007,UsovMelrose1995}.   
In this paper we consider the influence of
small scale magnetic field on polar cap heating
by reverse positron current
with taking into account
positronium generation and its photoionization 
by thermal photons from star surface.
Pulsar is considered in inner gap model with
free electron emission from neutron star surface.
We assume that pulsar diode is in stationary state
and take into account only positron generation due to
curvature radiation of primary electrons
and its absorption in magnetic field.

\begin{figure}[h]
\begin{minipage}{14pc}
\includegraphics[width=14pc]{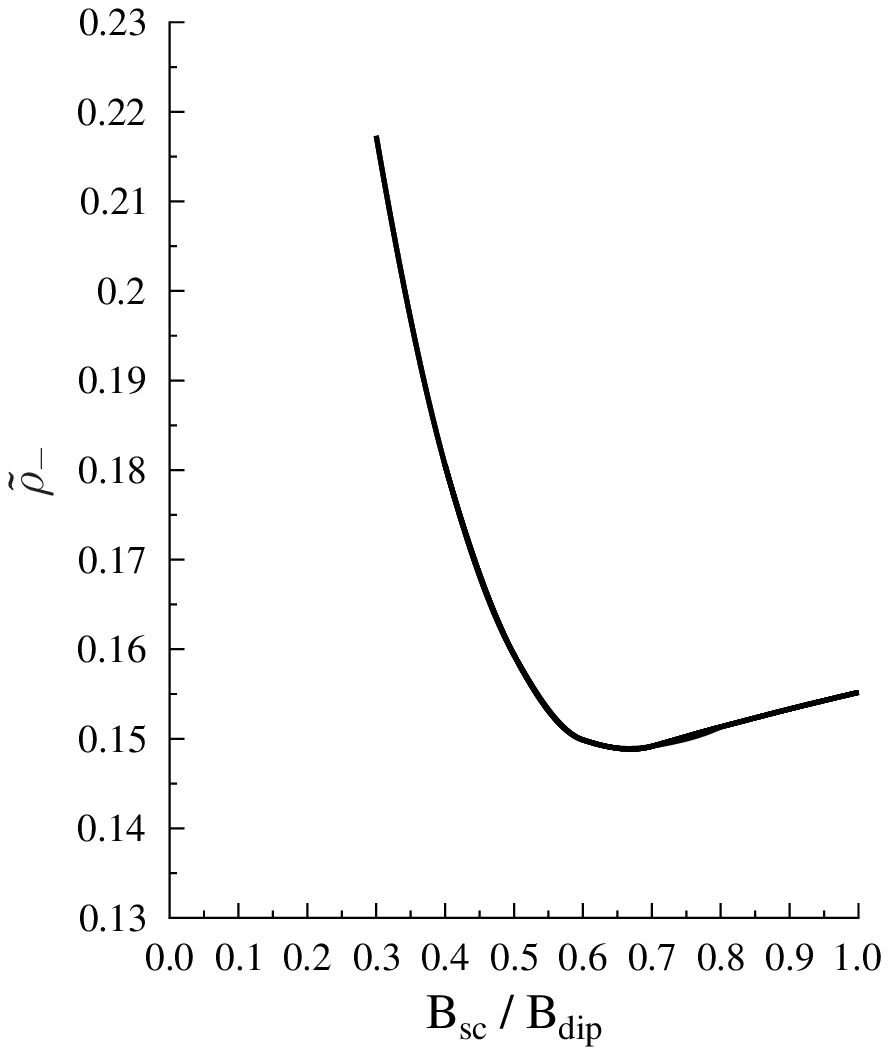}
\end{minipage}
\begin{minipage}{14pc}
\includegraphics[width=14pc]{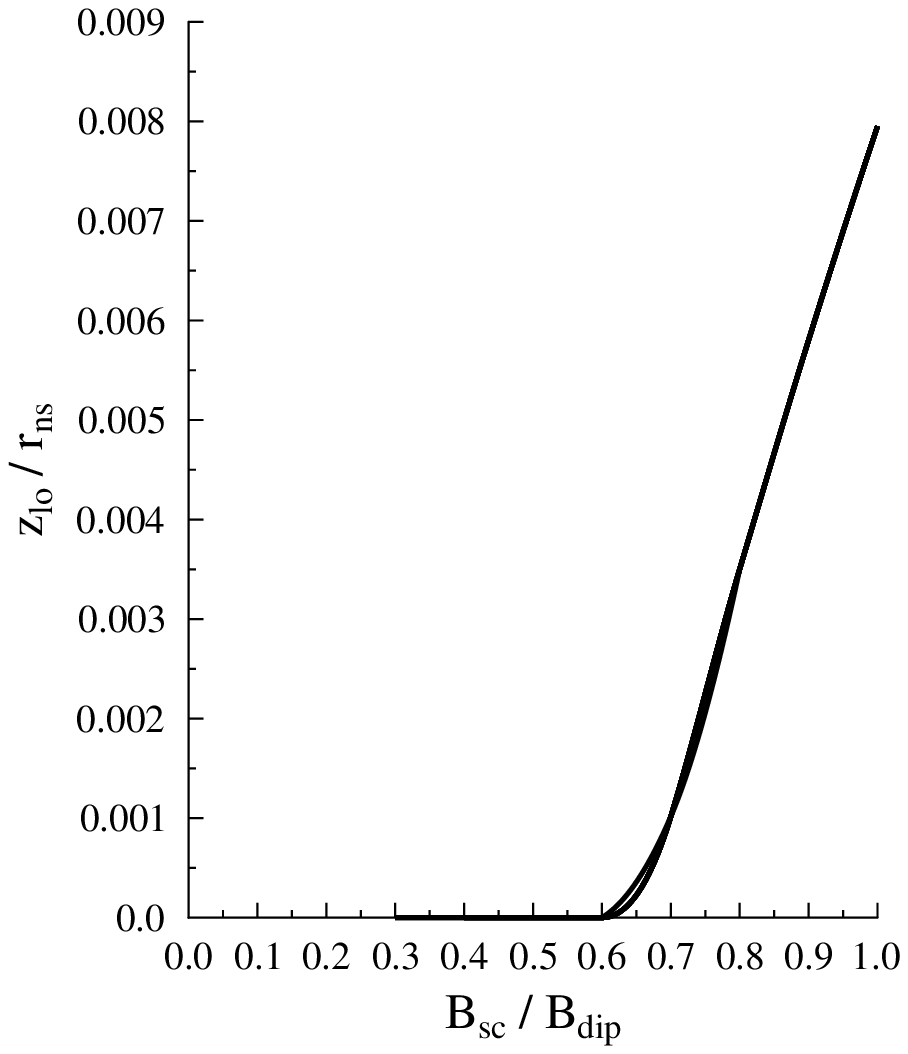}
\end{minipage}
\caption{\label{fig_Ae_z_lo}
The dependence of the primary electron density $\tilde{\rho}_{-}$
(in units $\frac{\Omega B}{2 \pi c}$) 
on small scale field strength $B_{sc}$ is shown
on the left panel. 
The dependence of the altitude $z_{lo}$ 
(in units $r_{ns}$) of diode lower plate (cathode) 
on small scale field strength $B_{sc}$ is shown
on the right panel.
}
\end{figure}
 
\begin{figure}[h]
\includegraphics[width=14pc]{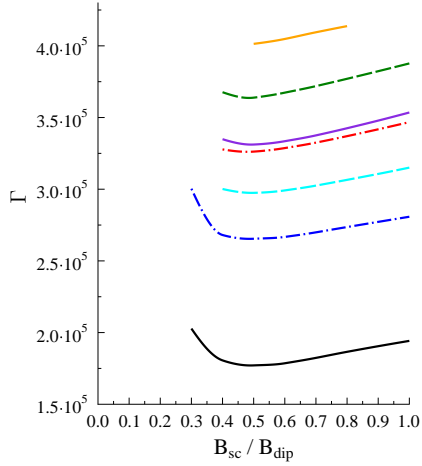}
\hspace{2pc}
\begin{minipage}[b]{14pc} 
\caption{\label{fig_G0}
The dependence of Lorentz factor 
$\Gamma = e \left. \Phi \right|_{z=z_{hi}} \, / mc^{2} $ 
of primary electrons
at central field line 
on small scale field strength $B_{sc}$ is shown.
Dot-dashed blue line corresponds to 
$T_{ns} = 3 \cdot 10^{5} \mbox{\ K}$ and $f=1$,
dashed cyan line corresponds to 
$T_{ns} = 3 \cdot 10^{5} \mbox{\ K}$ and $f=0.3$,
solid violet line corresponds to 
$T_{ns} = 3 \cdot 10^{5} \mbox{\ K}$ and $f=0.1$, 
dot-dashed red line corresponds to 
$T_{ns} = 10^{5} \mbox{\ K}$ and $f=1$,
dashed green line corresponds to 
$T_{ns} = 10^{5} \mbox{\ K}$ and $f=0.3$.
solid orange line corresponds to 
$T_{ns} = 1 \cdot 10^{5} \mbox{\ K}$ and $f=0.1$, 
The case $W_{0} = +\infty$ 
(all positroniums are photoionized immediately)
is shown by solid black line.
}
\end{minipage} 
\end{figure}

\begin{figure}[h]
\begin{minipage}{14pc}
\includegraphics[width=14pc]{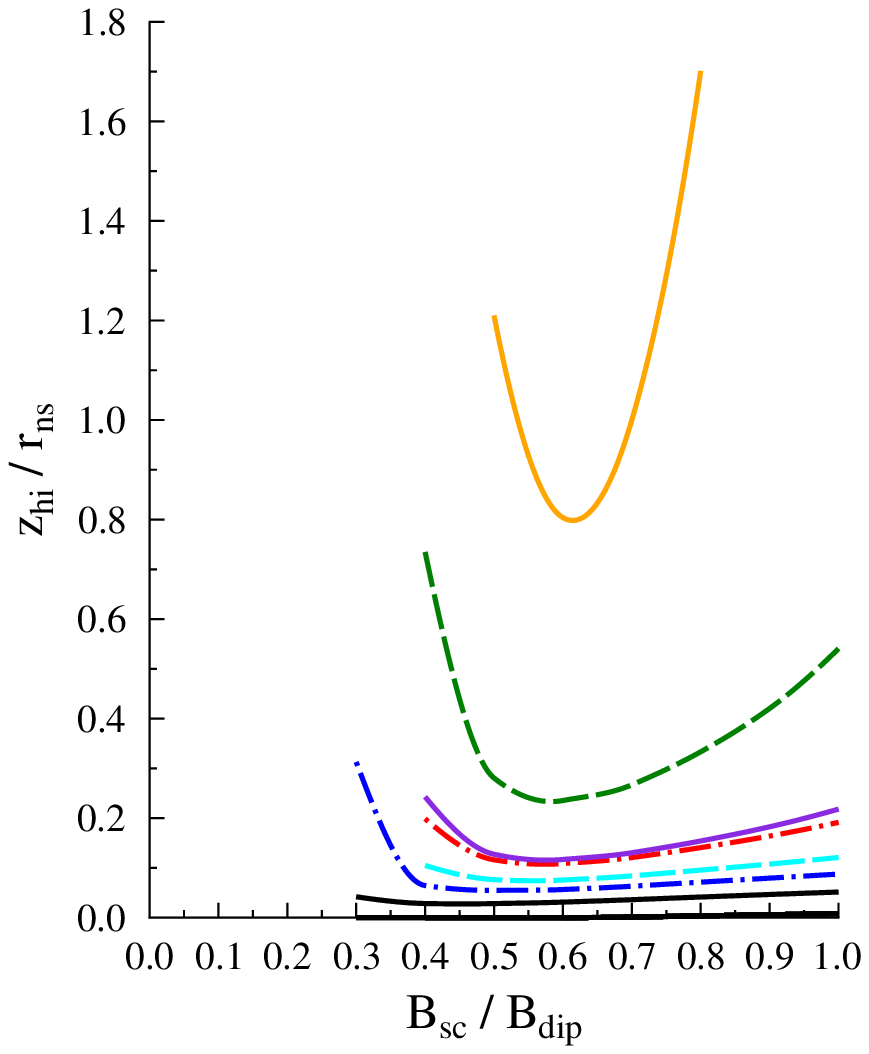}
\end{minipage}
\begin{minipage}{14pc}
\includegraphics[width=14pc]{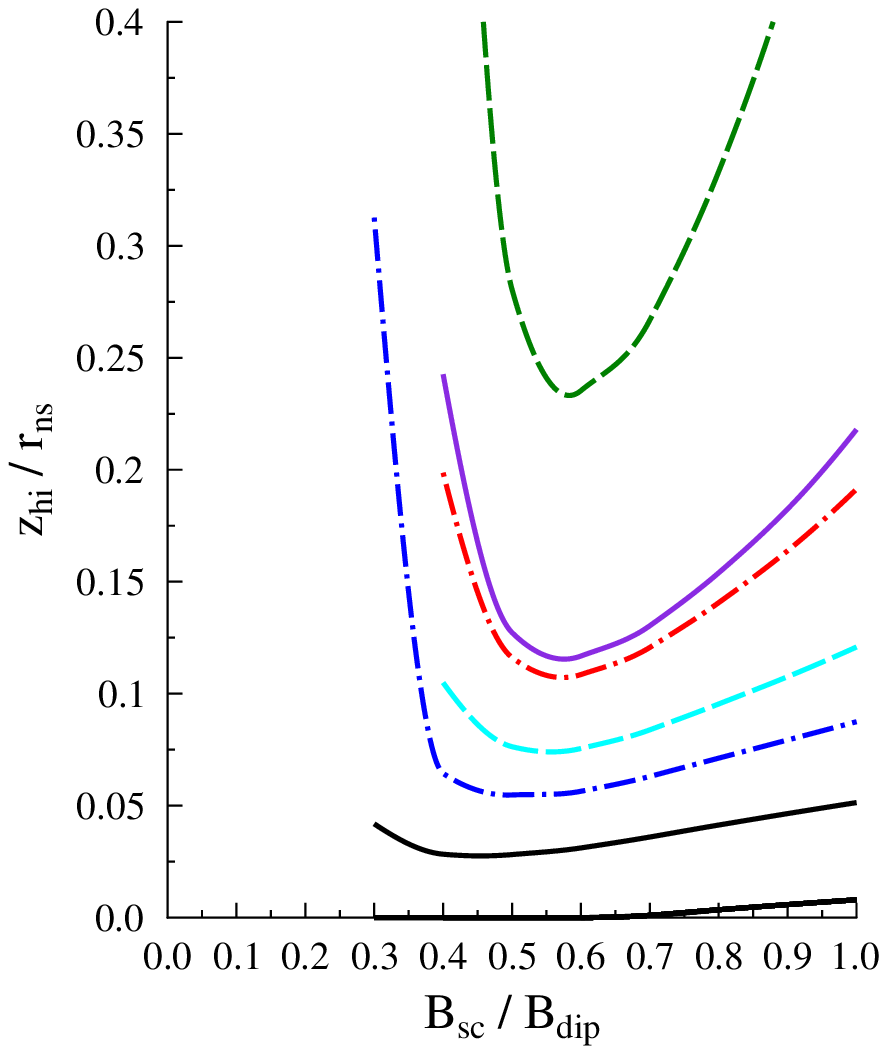}
\end{minipage}
\caption{\label{fig_z_hi}
The same as fig. \ref{fig_G0}, but 
the dependence of the altitude $z_{hi}$ 
(in units $r_{ns}$) of diode upper plate (anode) 
on small scale field strength $B_{sc}$ is shown.
The altitude of diode lower plate (cathode) $z_{lo}$ is
shown by black solid line on both graphs. 
Left and right graphs differ only in scale.
}
\end{figure}
 
\begin{figure}[h]
\begin{minipage}{14pc}
\includegraphics[width=14pc]{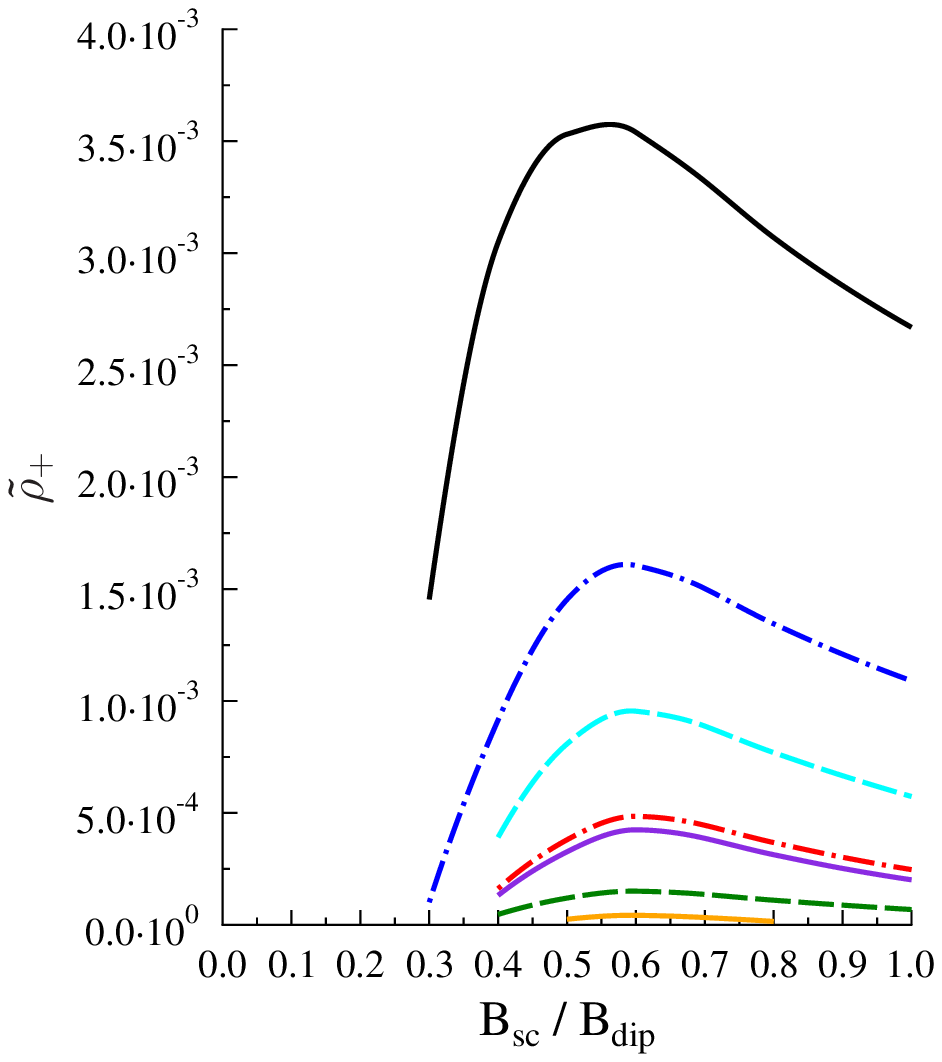}
\end{minipage}
\begin{minipage}{14pc}
\includegraphics[width=14pc]{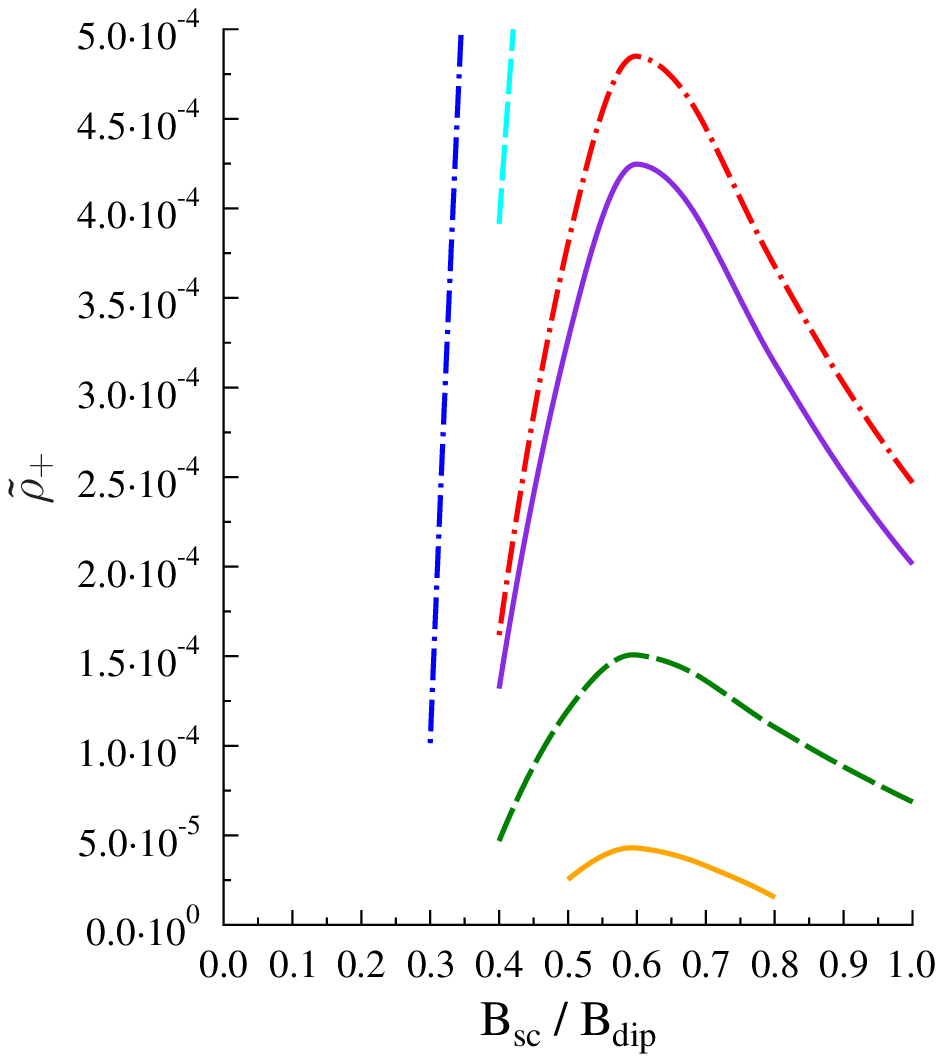}
\end{minipage}
\caption{\label{fig_AS_Ap}
The same as fig. \ref{fig_G0}, but 
the dependence of the reverse positron current $\tilde{\rho}_{+}$ 
(in units $\frac{\Omega B}{2 \pi c}$)
calculated with rapid screening model
on small scale field strength $B_{sc}$ is shown.
Left and right graphs differ only in scale.
}
\end{figure}

\begin{figure}[h]
\begin{minipage}{14pc}
\includegraphics[width=14pc]{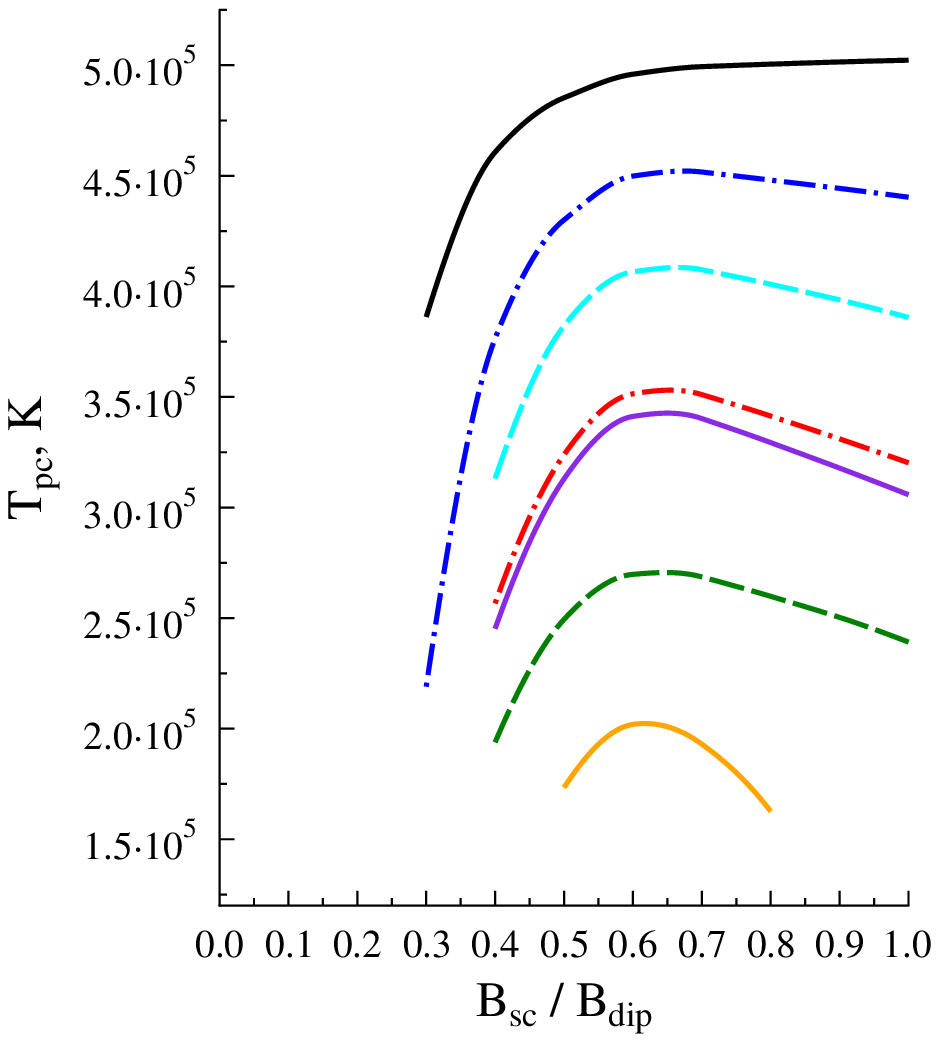}
\end{minipage}
\begin{minipage}{14pc}
\includegraphics[width=14pc]{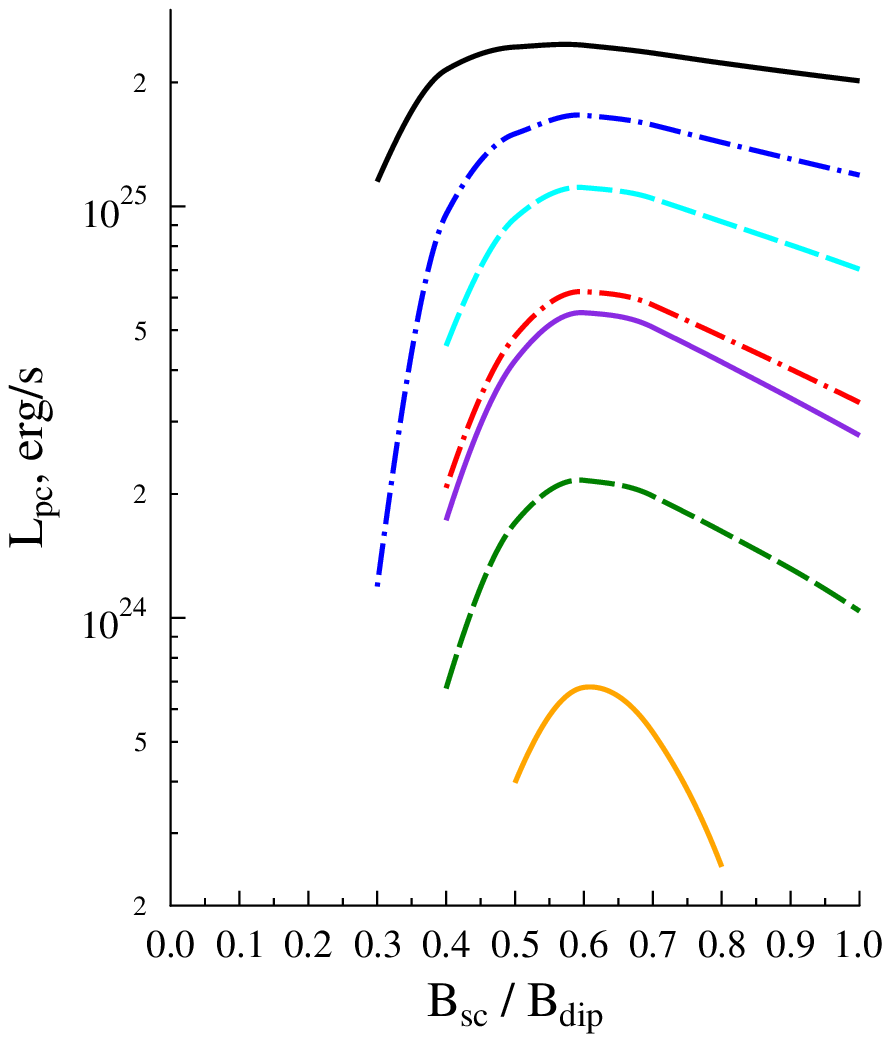}
\end{minipage}
\caption{\label{fig_AS_Lpc}
The same as fig. \ref{fig_G0}, but 
the dependence of the input of reverse positron heating to
polar cap surface temperature $T_{pc}$ at polar cap center
on small scale field strength $B_{sc}$ is shown
on the left panel. 
The dependence of corresponding polar cap luminosity $L_{pc}$
on small scale field strength $B_{sc}$ is shown
on the right panel.
Both panels correspond to rapid screening model. 
}
\end{figure}

\begin{figure}[h]
\begin{minipage}{14pc}
\includegraphics[width=14pc]{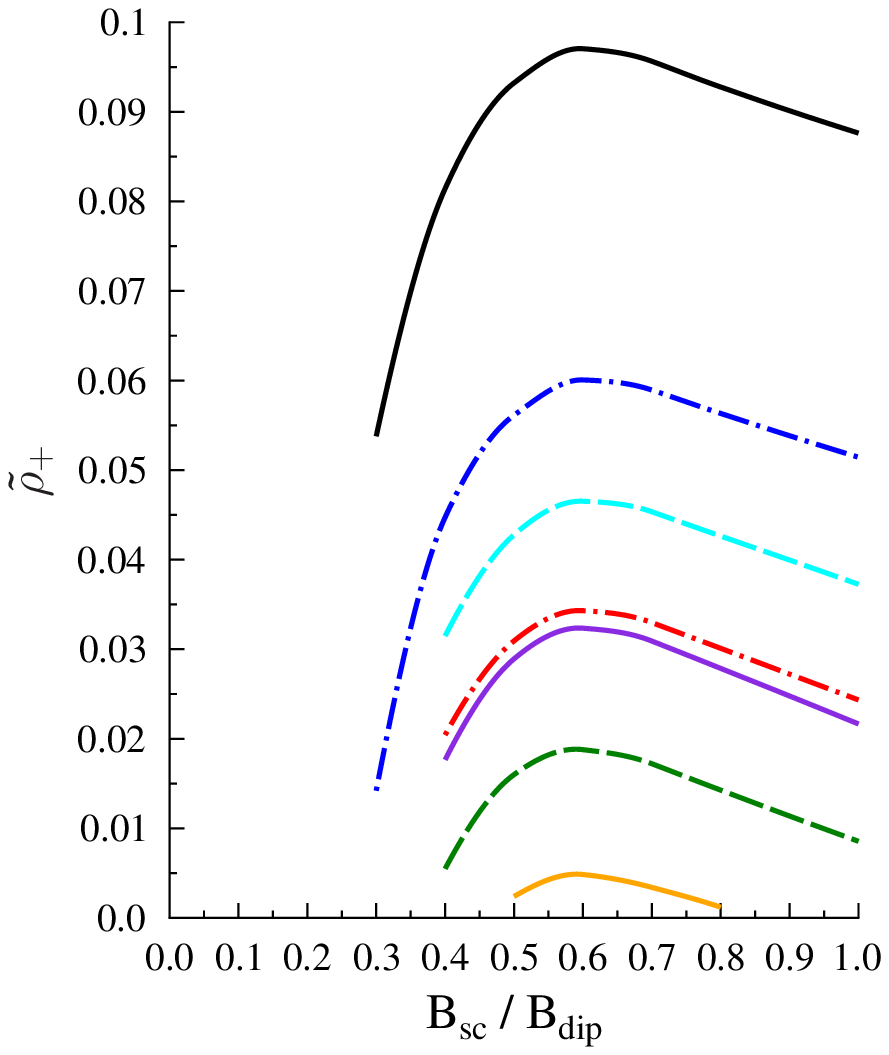}
\end{minipage}
\begin{minipage}{14pc}
\includegraphics[width=14pc]{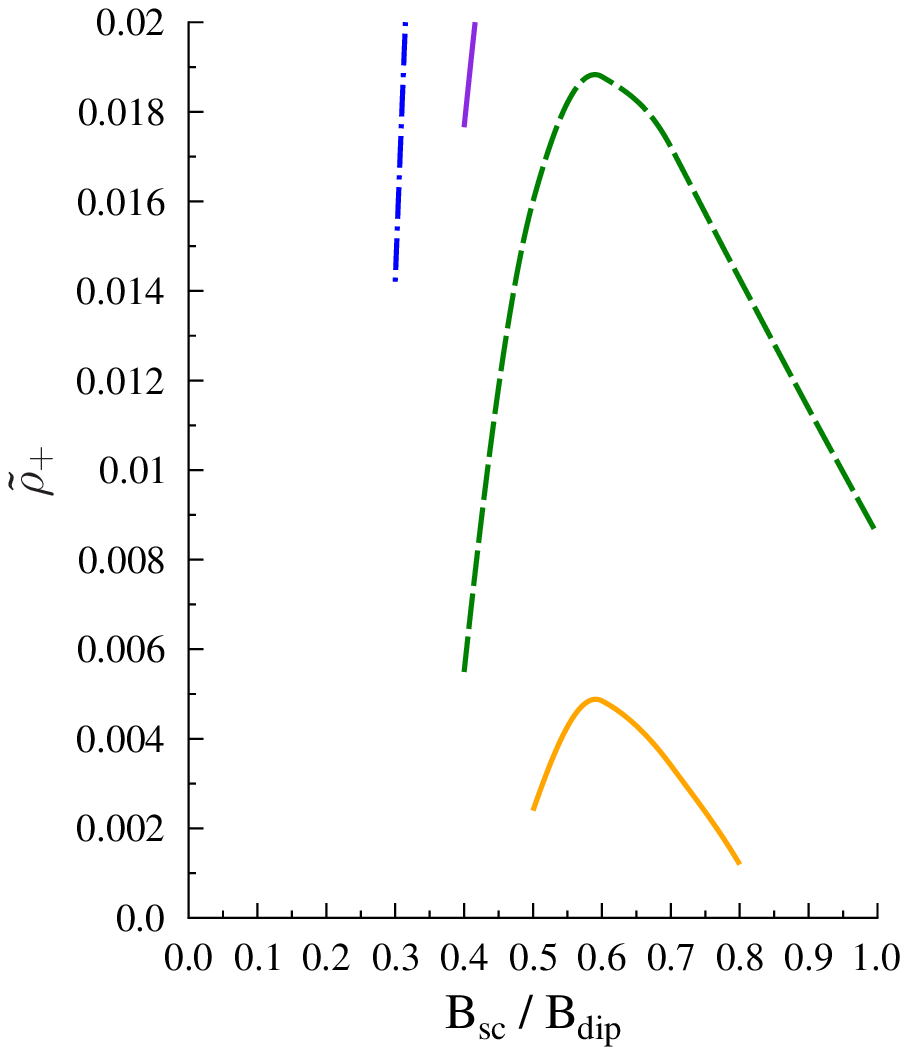}
\end{minipage}
\caption{\label{fig_MH_Ap}
The same as fig. \ref{fig_AS_Ap}, but
the case of gradually screening model is shown. 
Left and right graphs differ only in scale.
}
\end{figure}

\begin{figure}[h]
\begin{minipage}{14pc}
\includegraphics[width=14pc]{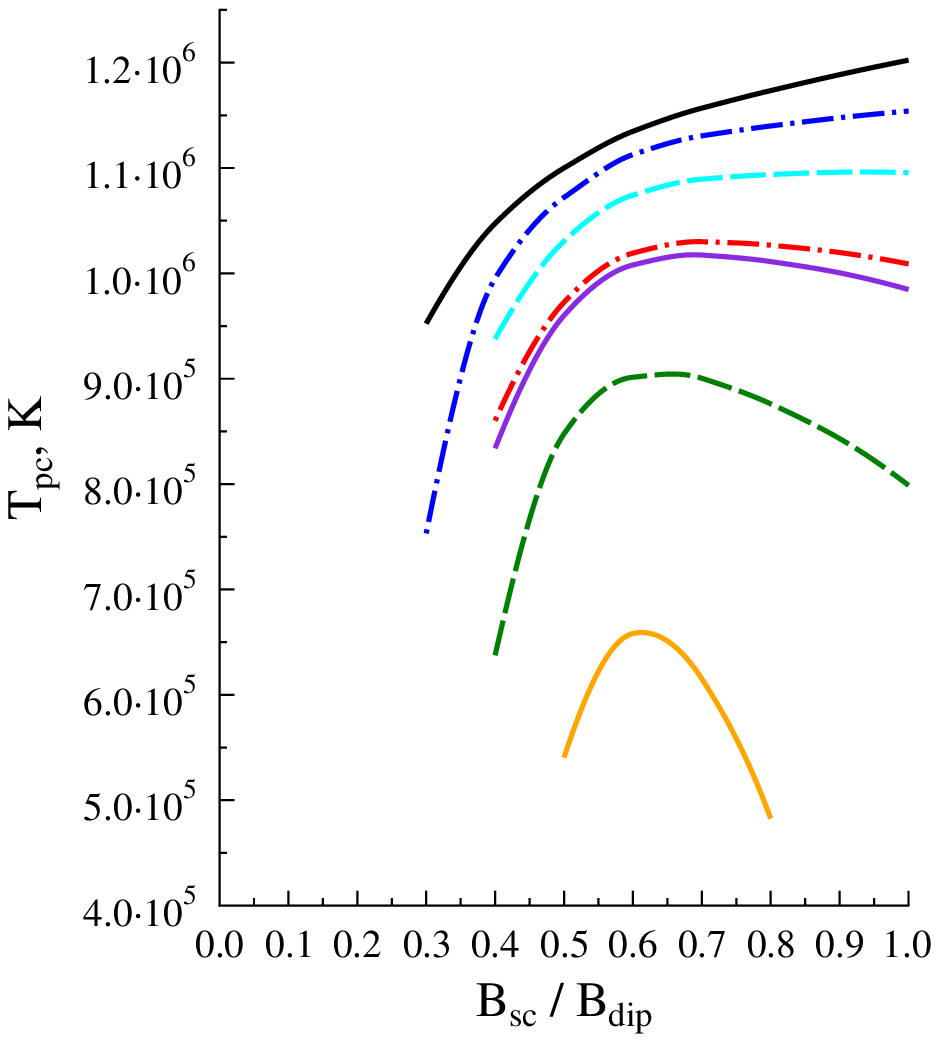}
\end{minipage}
\begin{minipage}{14pc}
\includegraphics[width=14pc]{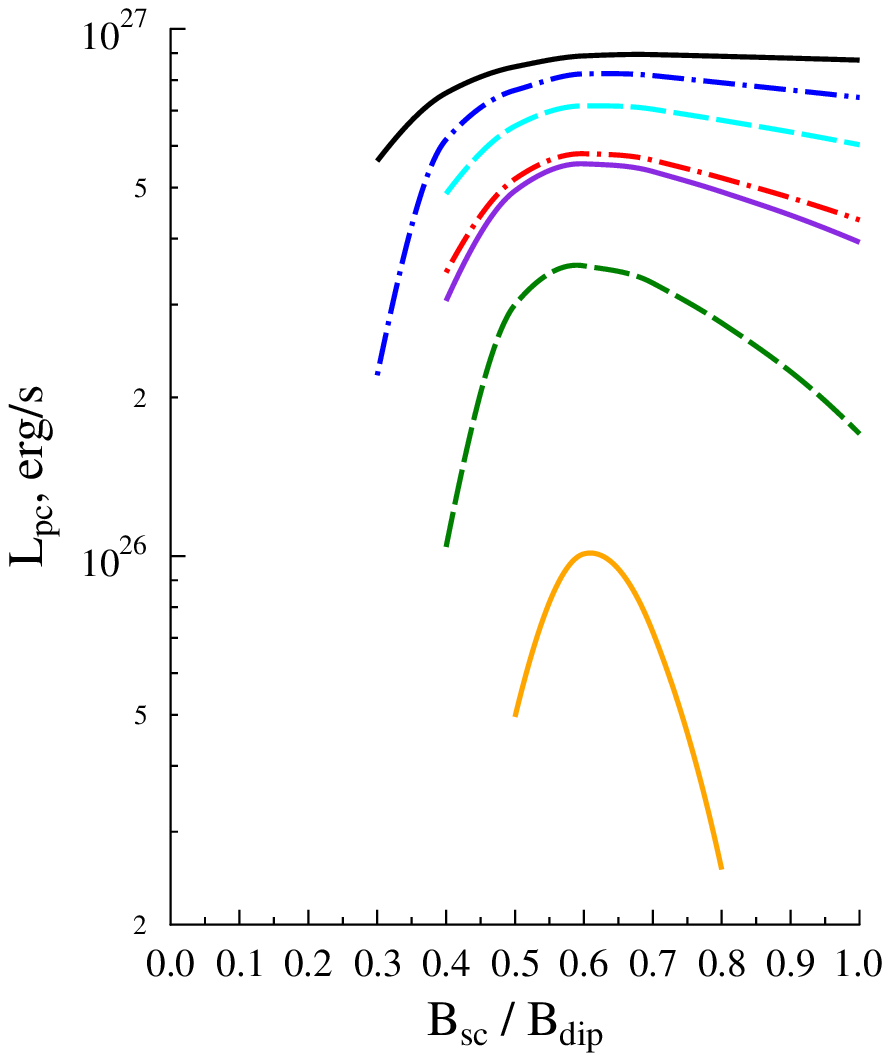}
\end{minipage}
\caption{\label{fig_MH_Lpc}
The same as fig. \ref{fig_AS_Lpc}, but 
the case of gradually screening model is shown.
}
\end{figure}

\begin{figure}[h]
\begin{minipage}{14pc}
\includegraphics[width=14pc]{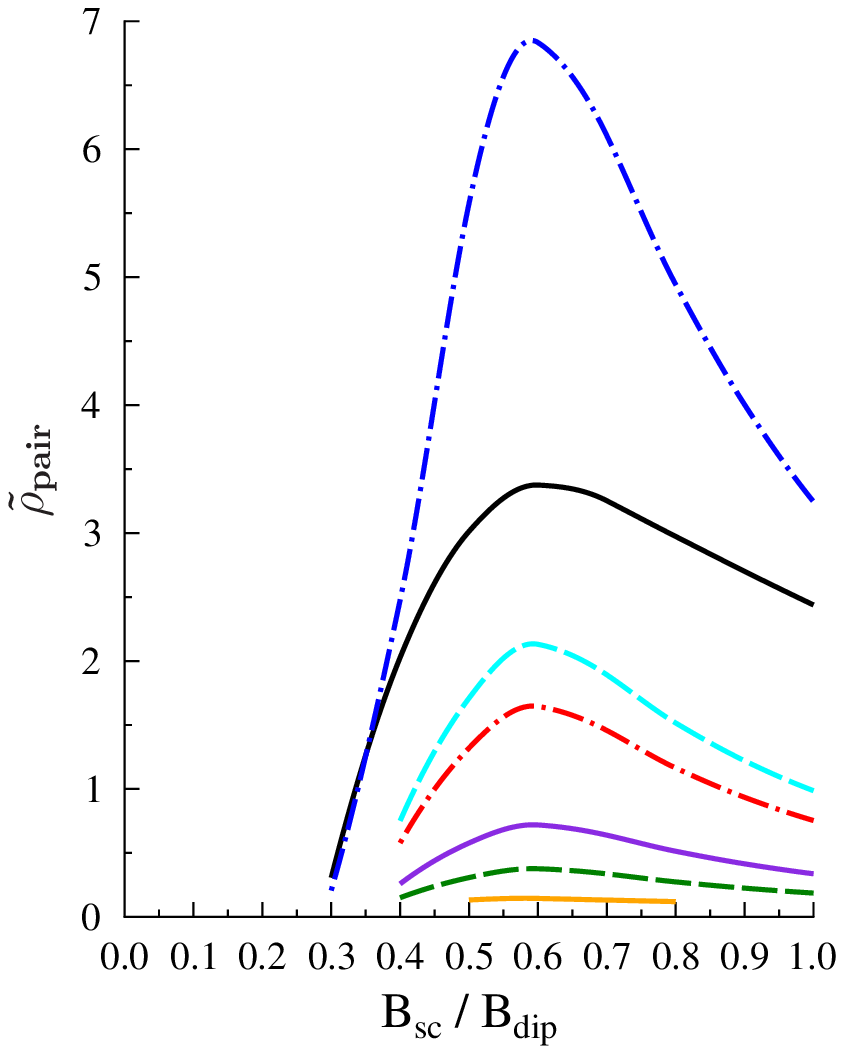}
\end{minipage}
\begin{minipage}{14pc}
\includegraphics[width=14pc]{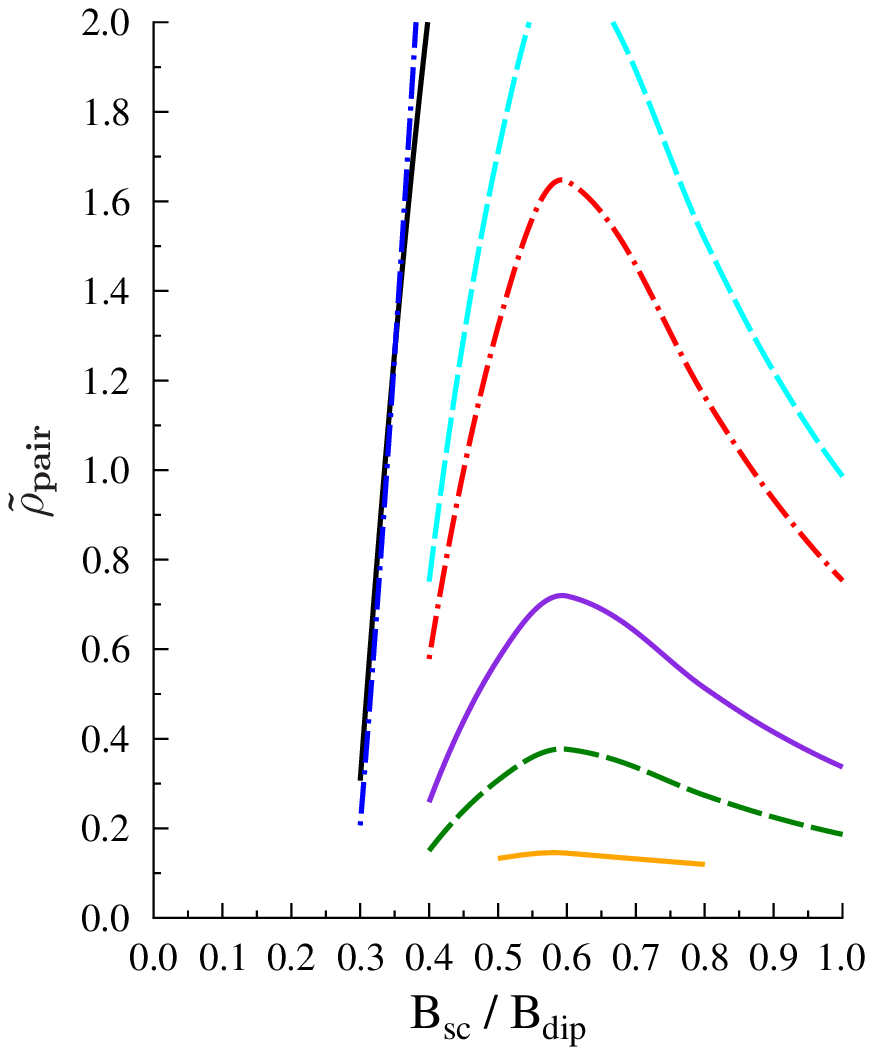}
\end{minipage}
\caption{\label{fig_Qtot}
The same as fig. \ref{fig_G0}, but 
the dependence of the total number of 
produced unbound or photoionized pairs $\tilde{\rho}_{pair}$
(in units $\frac{\Omega B}{2 \pi c}$)
on small scale field strength $B_{sc}$ is shown.
Left and right graphs differ only in scale.
}
\end{figure}

\section{Model}
Let the neutron star have a radius $r_{ns}$ and dipolar
magnetic moment $\vec{m}$ 
(its field at magnetic pole is $B_{dip} = 2m / r_{ns}^{3}$).
We assume also that a small-scale magnetic field 
with strength $B_{sc}$ and characteristic scale $\ell$
presents nearby the polar cap.
For simplicity we model small-scale magnetic field by
by additional magnetic moment $\vec{m}_{sc}$ 
locating in the polar region of the neutron star 
at depth $\ell $
\cite{Mitra2002,Kantor2003,Szary2013}: 
\begin{equation}
        \vec{B} = 
            \frac{
                     3 \vec{r} \left( \vec{r} \cdot \vec{m} \right)
                   - \vec{m} r^{2}
                 }
                 {
                     r^{5}
                 }
          + \frac{
                     3 \vec{\rho} \left( \vec{\rho} \cdot \vec{m}_{sc} \right)
                   - \vec{m}_{1} \rho^{2}
                 }
                 {
                     \rho^{5}
                 }  
\end{equation}
where $\vec{\rho} = \vec{r} - (r_{ns} - \ell) \vec{e}_{z}$,
$\vec{m} = m \vec{e}_{z}$,
$B_{sc} = m_{sc} / \ell^{3}$ -- small-scale field strength
at (dipolar) magnetic pole.
For simplicity we suppose that the vector $\vec{m}_{sc}$ lies
parallel to surface (and $\vec{m}_{sc} \cdot \vec{m} = 0$) 
in the plane containing $\vec{m}$ and $\vec{\Omega}$
and is directed "along"\  $\vec{\Omega}$,
see fig. \ref{fig_diode}.
Hence surface small-scale magnetic field is directed
"opposite"\  pulsar rotation velocity $\vec{\Omega}$,
i.e. $\psi_{\Omega} = 0$, see fig. \ref{fig_angles}.
Also we assume that inclination angle $\chi$
is equal to $\chi = 60^{\circ}$.

%
%
%
We consider only the case of inner gap \cite{Shibata1991}
and assume that the inner gap occupies the entire pulsar tube cross section 
and resides as low as possible.
Let us denote the altitudes of  inner gap lower plate (cathode)
and upper plate (anode) 
by $z_{lo}$ and  is $z_{hi}$ respectively,
see fig. \ref{fig_diode}.
In most cases the inner gap resides 
exactly on neutron star surface ($z_{lo} = 0$), 
see \cite{Polyakova2009} for details.
We suppose that the inner gap is stationary  and
operates in the regime of charge limited steady flow \cite{Arons1977}.
%
Hence in the reference frame rotating with the star 
we can write,
see \cite{Barsukov2016} for details:
\begin{equation}
         \Delta \Phi = - 4\pi ( \rho - \rho_{GJ} )
         \mbox{,\ \ }
         \vec{E} = - \vec{\nabla}\Phi 
\end{equation}
\begin{equation}
\left. E_{||} \right|_{z=z_{lo}} = 0 
\mbox{ \ and \ }
\left. E_{||} \right|_{z=z_{hi}} = 0  
\end{equation}
\begin{equation}
\left. \Phi \right|_{z=z_{lo}} = 0
\mbox{ \ and \ }
\left. \Phi \right|_{side} = 0
\end{equation}
where $z$ is the altitude above star surface, 
$\Phi$ is electrostatic potential,
$\left. \Phi \right|_{side}$ is its value at pulsar tube boundary, 
$\rho_{GJ} = \frac{\Omega B}{2\pi c} \tilde{\rho}_{GJ}$ 
is Goldreich-Julian density \cite{GoldreichJulian1969},  
$\rho = \frac{\Omega B}{2\pi c} \tilde{\rho}$ is total charge density,
$\tilde{\rho} = \tilde{\rho}_{-} + \tilde{\rho}_{+}$,
$\tilde{\rho}$, $\tilde{\rho}_{-}$, $\tilde{\rho}_{+}$
are total charge density, electron and positron densities
in units $\frac{\Omega B}{2\pi c}$ correspondingly.
%
We assume that inside pulsar diode
the particles 
move along field lines $\vec{v} \parallel \vec{B}$
with relativistic velocity $v \approx c$.
So continuity equation 
${\mathrm{div}}\left( \rho_{\pm} \vec{v} \right) = 0$
may be rewritten as  
$(\vec{B} \cdot \vec{\nabla}) \tilde{\rho}_{\pm} = 0$
and hence densities $\tilde{\rho}_{\pm}$ are constant along field lines.
Also it is worth to note that without frame dragging
$\tilde{\rho}_{GJ}(\vec{x}) \approx - \cos\tilde{\chi}$,
where $\tilde{\chi}$ is the angle between
field vector $\vec{B}(\vec{x})$ at point $\vec{x}$ 
and angular velocity vector $\vec{\Omega}$.

For simplicity we take into account 
the generation of electron-positron pairs
only by curvature radiation of primary electrons in magnetic field. 
We also take into account the generation of pairs in bound state (positronium).
For simplicity  we assume that the probability $P_{b}$
that a pair is created in bound state is defined as follows
\cite{UsovMelrose1995}:
\begin{itemize}
\item[]
$P_{b}=0$ if $B < B_{low}$ (no positroniums are created),
\item[]
$P_{b} = (B-B_{low}) / (B_{high}-B_{low})$
if $B_{low} \leq B \leq B_{high}$ and
\item[]
$P_{b} = 1$ if $B > B_{high}$
(all pairs are created in bound state),
\end{itemize}
where $B$ is magnetic strength at point of  pair  creation,
$B_{low}=0.04 \, B_{cr}$, $B_{high}=0.15 \, B_{cr}$,
$B_{cr} \approx 4.41 \cdot 10^{13} G$ \cite{UsovMelrose1995}.
In order to simplify the calculation we assume 
that the pair generation and its properties do not depend on
photon polarization.
%
However, we take into account the photoionization of positronium  
by thermal photons from hot polar cap.
The photoionization rate
is estimated by formula  \cite{UsovMelrose1995} 
\begin{equation}
\frac{dN}{dt}(\vec{x})
  = W_{0} \left( \frac{10^{2}}{\Gamma} \right)^{3}
                    \left( \frac{ T_{ns} }{ 10^{6} K } \right)^{2}
                    (1 - \cos\theta_{ns}),
\end{equation} 
where $\Gamma$ is positronium Lorentz factor, 
$T_{ns}$ is neutron star surface temperature,
$\theta_{ns}$ is angular radius of neutron star
at point $\vec{x}$,
$W_{0} = 6 \cdot 10^{5} \mbox{s}^{-1}$ \cite{UsovMelrose1995}.  
Due to small polar cap size we neglect
positronium photoionization by thermal photons
from hot polar cap.
In this paper we does not take into account photon splitting
and positronium decay. 
In order to crude estimate the effect of these processes
we assume that $(1-f)$ part of positroniums immediately decays
after creation and $f$ part of positroniums does not decay at all.
%
%
%
%
%
%
%

The calculations of reverse positron current 
are performed in two 
models based on extreme assumptions 
about the rate of parallel electric field
$E_{||} = (\vec{E} \cdot \vec{B}) / B$
screening: 
the model of rapid screening \cite{Arons1979_Pairs_production}
according to which the electron-positron plasma  screens
parallel electric field 
almost immediately
and the model of gradually screening 
\cite{HardingMuslimov2001,Lyubarskii1992},
which allows the parallel electric field penetrates deep 
into electron-positron plasma,
see details of calculation in \cite{Barsukov2016}.
For simplicity we assume that anode altitude $z_{hi}$ 
is determined by equation
\begin{equation}
\left. \tilde{\rho}_{pair} \right|_{z=z_{hi}} = 
{\mathrm{max}}
         \left( \frac{1}{10} , 
                \tilde{\rho}^{r}_{+}
         \right)
\end{equation}
where $\tilde{\rho}_{pair}$ is number unbound electron-positron pairs
generated at central field line,
$\tilde{\rho}^{r}_{+}$ is reverse positron density
calculated according rapid screening model.
%
%
The input of reverse positron heating to
polar cap temperature $T_{pc}$ is estimated as
\begin{equation}
\sigma_{B} T_{pc}^{4} =
\left. \Phi \right|_{z=z_{hi}} \cdot 
\left. \frac{\Omega B}{2\pi} \right|_{z=0} \cdot
\tilde{\rho}_{+}
\end{equation}
where altitude $z=0$ corresponds to star surface,
$\sigma_{B}$ is Stefan-Boltzmann constant
and all values calculated at the same field line.
Polar cap luminosity due to reverse positron heating
is estimated as
\begin{equation}
L_{pc} = \int_{S_{pc}} \sigma_{B} T_{pc}^{4} \, dS
       \approx
         \left. \Phi \right|_{z=z_{hi}} \cdot
         \frac{\Omega B_{dip} }{2\pi} \cdot
         \tilde{\rho}_{+} \cdot 
         \pi \, \left( \theta_{0} r_{ns} \right)^{2}
\end{equation}
where we integrate over polar cap surface and
$\theta_{0} = \sqrt{ \Omega r_{ns} / c }$,
see \cite{Barsukov2016} for details.

\section{Results}
The dependence of primary electron current $\tilde{\rho}_{-}$ 
and  diode lower plate (cathode) altitude $z_{lo}$  
on strength of small scale magnetic field $B_{sc}$
is shown in fig. \ref{fig_Ae_z_lo}.
According to used model altitude $z_{lo}$ does not
depend on pair production at all \cite{Polyakova2009}.
And because of pulsar tube radius is very small 
$\theta_{0} r_{ns} \ll z_{hi}$ 
primary electron density $\tilde{\rho}_{-}$
also does not depend on pair production.
%
The dependence of energy of primary electrons and
diode upper plate (anode) altitude $z_{hi}$  
on strength of small scale magnetic field $B_{sc}$
are shown in fig. \ref{fig_G0} and fig. \ref{fig_z_hi}
correspondingly.
At $B_{sc} \lesssim 0.5 B_{dip}$ altitude $z_{hi}$
decreases with increasing $B_{sc}$ 
because of increasing total magnetic field strength
and, most importantly, due to increasing field line curvature.
Later according considered magnetic field model
cathode altitude $z_{lo}$ begins to increase.
Hence primary electrons are accelerated 
at larger altitude where field strength and
its curvature are less.
And consequently pair production becomes less effective
and anode altitude $z_{hi}$ grows with $B_{sc}$
at $B_{sc} \gtrsim 0.6 B_{dip}$.
And the increasing of the altitude $z_{hi}$ causes
an increase in the energy of primary electron.
The dependence of reverse positron current $\tilde{\rho}_{+}$
and the polar cap luminosity $L_{pc}$ caused by this current
on strength of small scale magnetic field $B_{sc}$ 
in case of rapid and gradually screening model 
are shown in figures \ref{fig_AS_Ap}-\ref{fig_MH_Lpc}.
The dependence of total number of 
produced unbound or photoionized pairs $\tilde{\rho}_{pair}$
on strength of small scale magnetic field $B_{sc}$ 
is shown in fig. \ref{fig_Qtot}.
It is worth to note that number of pairs $\tilde{\rho}_{pair}$ 
produced in case of $T_{ns} = 3 \cdot 10^{5} \mbox{\ K}$ and $f=1$
is larger than in case of $W_{0} = +\infty$.

\section{Discussion} 
We consider inner gap model with 
stationary space charge limited flow
in J0250+5854 pulsar
and show that this pulsar may lye upper than pulsar "death line"\ 
in case of two assumption:
the presence of surface magnetic field
with very small characteristic scale $\ell \approx 500 \mbox{ m}$
and neutron star surface temperature 
$T_{ns} \sim (1-3) \cdot 10^{5} \mbox{ K}$.
Main problem is that the pulsar is very old
$\tau=13.7 \cdot 10^{6} \mbox{ years}$.
Hence it is difficult to explain 
why field with so small scale has survived
and why star is so hot.

It is worth to note that radiopulsar B0950+08 has 
spin down age $\tau = 17.5 \cdot 10^{6} \mbox{\ years}$
and star surface temperature 
$T_{ns} \sim (1-3) \cdot 10^{5} \mbox{ K}$
\cite{Pavlov2017}.
A such temperature may be related to
internal heating mechanisms 
like rotochemical heating and heating due to vortex friction
\cite{Guillot2019}.
We also may speculate that magnetic field decay event with Hall cascade
has occur not so long ago in this pulsar \cite{Igoshev2018}.
Hence small scale magnetic field may be generated during Hall cascades
and accompanying field decay may heat up the star.

In the paper we does not take into account 
the photon polarization and, consequently,
we can not estimate input of photon splitting effect 
and positronium decay,
see, for example, \cite{Baring2001,Sobyanin2007,Timokhin2019}.
Hence, our conclusion that magnetic field with 
characteristic scale $\ell \approx 500 \mbox{ m}$ is enough
to explain radio radiation of the pulsar 
may be too optimistic.
But we hope that field with $\ell \approx 300 \mbox{ m}$
would be enough.
Also it is worth to note that
we take into account only curvature radiation of primary electrons
and resonant compton scattering may give 
a similar quantity of pairs \cite{Muslimov2001}.

Our choice of inclination angle $\chi = 60^{\circ}$
does not motivated by anything.
Although we find that in considered field configuration
the pulsar lye down "pulsar death"\  line in case of
$\chi = 0^{\circ}$ and $\chi = 30^{\circ}$.
But we guess that it is only artifact of used small scale field model.

\ack
We sincerely thank 
A.I. Tsygan, O.A. Goglichidze, K.Yu. Kraav, V.M. Kontorovich,
D.A. Rumyantsev, D.N. Sobyanin, I.F. Malov and V.A. Urpin
for help, comments and usefull discussions.
We thank V.S. Beskin for pointing out the other explanation
of the existence of J0250+5854 radio radiation
and usefull discussion about this pulsar.

\end{document}